\documentstyle[aps, epsf]{revtex}
\begin{document}
\wideabs{
\title{Simulations of a single membrane between two walls\protect\\
using a Monte Carlo method}
\author{Nikolai Gouliaev and John F. Nagle}
\address{Department of Physics
\protect\\Carnegie Mellon University , Pittsburgh, PA 15213}
\date{\today}
\maketitle

\begin{abstract}
Quantitative theory of interbilayer interactions is essential to
interpret x-ray scattering data and to elucidate these
interactions for biologically relevant systems.  For this purpose
Monte Carlo simulations have been performed to obtain pressure $P$ and
positional fluctuations $\sigma$.  
A new method, called Fourier Monte-Carlo (FMC), that is based 
on a Fourier representation of the displacement field, is developed and
its superiority over the standard method is demonstrated. 
The FMC method is applied to simulating a single membrane
between two hard walls, which models a stack of lipid bilayer
membranes with non-harmonic interactions. Finite size scaling is
demonstrated and used to obtain accurate values for $P$ and $\sigma$
in the limit 
of a large continuous membrane. The results are 
compared with perturbation theory approximations, and numerical
differences are found in the non-harmonic case. Therefore, the FMC
method, rather than the approximations, should be used for establishing the
connection between model potentials and observable quantities, as well
as for pure modeling purposes.
\end{abstract}
}

\section{Introduction}
Recent research on lipid bilayers\cite{HP97} 
has contributed to the important biological physics goal
of understanding and quantifying the interactions between membranes
by providing high resolution x-ray scattering data.  From these
data the magnitude of fluctuations in the water spacing between
membranes in multilamellar stacks is obtained.  This enables
extraction of the functional form of the
fluctuational forces, originally proposed by Helfrich \cite{Hel78}
for the case of hard confinement.  For systems with large water
spacings, the Helfrich theory has been experimentally confirmed \cite{Saf89}.
For lecithin lipid bilayers, however, the water spacing is limited
to $20\AA$ or less.  For this important biological model system, our data 
show that a theory of soft confinement with a different functional
form is necessary; this is not surprising because interbilayer interactions
consist of more than hard-wall, i.e., steric interactions.

The theory of soft confinement is even more difficult than the
original Helfrich theory of hard confinement.  Progress has
been made by modeling the stack of interacting flexible membranes
by just one flexible membrane between two rigid walls\cite{SO86,PP92}. 
Even with this simplification, however, the theory involves an
uncontrolled approximation using first order perturbation theory and
a self-consistency condition 
in order that the interbilayer interaction may be approximated
by a harmonic potential\cite{PP92}.  We have obtained inconsistent
results when applying this theory to our data (unpublished).
Possible reasons are
(i) the theory is quantitatively inaccurate or (ii) the single membrane 
model is too simple.  The immediate motivation for this paper is
to test possibility (i). 

In order to obtain accurate results for a system with realistic
non-harmonic potentials, we use Monte-Carlo (MC) simulations.
The particular MC method developed in this paper will be called
the FMC method because it uses the Fourier representation for the
displacement of the membrane rather than the customary pointwise
representation, which will be called the PMC method.
The main advantage of the FMC method is that the optimal step sizes do  
not decrease as more and more amplitudes are considered. In contrast,
in PMC simulations, the optimal step sizes decrease as the inverse of the
density of points in one dimension, because the bending energy becomes
large when single particle excursions make the membrane rough. Because of this,
relatively large moves of the whole membrane are possible with the
FMC method, but not the PMC method.
This produces rapid sampling of the whole accessible phase space, while
respecting the membrane's smoothness. The resulting time series have
moderate auto-correlation times \cite{FN1} that do not increase substantially
as the membrane gets larger and/or more amplitudes are taken into
account. Even though each Monte Carlo step takes longer,
FMC still outperforms PMC by
a wide margin. It then becomes possible to carry out substantial
simulations on a standalone workstation rather than a
supercomputer\cite{LipowskyZielinska}
and to obtain accurate results for a single membrane
subject to realistic potentials with walls, and even for a stack of
such membranes (to be described in a future paper) \cite{FNx}.

Section \ref{DefinitionOfTheSystem} defines the
membrane model and the physical quantities simulated in the paper. 
Section \ref{Details} describes the FMC method and also gives some
important details that are used to speed up the code.
In Section \ref{HarmonicPotentialAndScaling} the method is tested
on an exactly solvable model, namely, one that has only harmonic
interactions with the walls.  This test also allows examination of
the system properties and the convergence of FMC results for an
infinitely large, continuous membrane.  In section \ref{RealPot} the FMC
method is applied to a single membrane with realistic,
non-harmonic interactions with the walls.
Section \ref{ComparisonWithReal-Space} makes a detailed comparison of
the FMC method and the standard PMC method. This
section shows that the FMC method not only converges faster to average
values for continuous membranes, but also gives smaller stochastic errors.
Finally, section \ref{AccurateResults} compares simulation results
with those obtained using the analytic first-order theory
of Podgornik and Parsegian\cite{PP92} and from experiment\cite{HP97}.
  
\section{Single Membrane System}
\label{DefinitionOfTheSystem}
\vspace{-0.3in}
\begin{figure}[h]
\begin{center}
\leavevmode
\epsfxsize 7.0cm
\epsffile{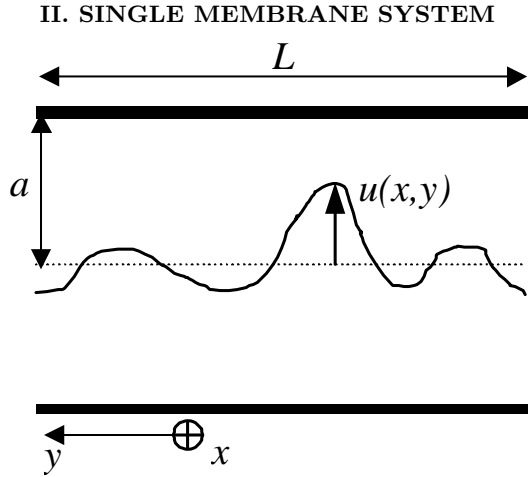}
\caption{A fluctuating single membrane, constrained between two hard walls.}
\label{walls}
\end{center}
\end{figure}
At the atomic scale a lipid membrane is composed of
complex lipid molecules and many simulations are performed at this scale
\cite{Pastor1,Berko1,Tu1}.  However, for modeling
the structure factor for low angle x-ray scattering (in contrast
to modeling the form factor), it is customary and appropriate
\cite{Cai72,Als80,Holyst91,Zhang94} to model the membrane as an
infinitely thin flexible sheet as shown in Fig.\ref{walls}.

The membrane undulates with instantaneous fluctuations in the
$z$-direction, given by
$u(x,y)$, subject to periodic boundary conditions.
The model energy $W$ is a sum of bending energy with
a bending modulus $K_c$ and an energy of interaction with the walls,
\begin{equation}
W = \frac{K_c}{2} \int ({\Delta}u)^2 dx\ dy + \int w_a(u) dx\ dy.
\label{RealHamiltonianOneMembrane}
\end{equation}
Since each wall is a surrogate for a neighboring membrane in a stack,
and since it is desired to obtain physical properties per membrane,
the interaction potential is given by the average 
$w_a(u) = (V(a+u) + V(a-u)) / 2$ of the interactions $V$ with each
wall and the corresponding volume of the system per membrane is then
$aL^2$.  For a separation $z$ between a wall and the membrane the
interaction potential will be based on the standard form
\begin{equation}
V(z) = A\lambda e^{-z/\lambda} - \frac{H}{12{\pi}z^2},
\label{V(z)}
\end{equation}
where the first term on the right hand side is a repulsive hydration potential
\cite{PP92} and the last term is an approximate, attractive van der
Waals potential. 
The divergence in the van der Waals potential as $z{\rightarrow}0$
in Eq.(\ref{V(z)}) is quite artificial; physically, it is masked by
stronger steric repulsions at small $z$ \cite{McI87a}.
This is corrected in this paper by including only a finite
number of terms $m_{max}$ in a power series expansion of $1/z^2$ about $u=0$.
It is shown later that a wide range of $m_{max}$ give nearly the same
result, so $m_{max}$ is not a critical parameter and
power series suffice to represent the van der Waals potential
satisfactorily for the most probable values of $z$ but avoid including
artificial traps near the walls. Other forms besides
Eq.(\ref{V(z)}) can be treated as well.

The first important quantity, obtained directly from the 
simulation, is the mean square fluctuation ${\sigma}^2$ in the water spacing.
In Fig.\ref{walls}, $\sigma^2=\overline{u^2(x,y)}$, where the average
is over both space and time.  The second
physical quantity is the pressure $P$ that 
must be exerted on the walls to maintain the average water spacing $a$.
The pressure is a sum of two 
components: $P_1$, caused by collisions and equal to a temporal
average of a delta-function-like instantaneous pressure, and $P_2$, which
is due to non-contact interactions with the walls, and that varies smoothly
with time and position. A virial theorem argument can be
used to compute $P_1$. The general result is
\begin{eqnarray}
P &=& \left[\frac{N^2 k_B T - 2\bar{U}}{2aL^2} - \frac{1}{2aL^2} \overline{\int u\frac{{\partial}w}{{\partial}u} dx\ dy}~\right] -\nonumber\\
& & \overline{\frac{{\partial}w(u, a)}{{\partial}a}},
\label{PSingleFinalResult}
\end{eqnarray}
where $P_1$ is the term in square brackets.
The relative importance of $P_1$ and $P_2$ depends on the
potential. If the potential is completely steric (hard wall), then
$P_2=0$. However, we have found that for the more realistic potentials
considered in this paper $P_1$ is very small compared to $P_2$ because
there are very few hard collisions.  

\section{Fourier Monte Carlo (FMC) Method}
\label{Details}
The membrane displacement $u(x,y)$ is represented by its Fourier
amplitudes $u(\vec{Q})$, where $\vec{Q} = (2{\pi}m/L, 2{\pi}n/L)$,
$N$ is the total number of modes in each dimension and 
$-N/2+1{\leq}m,n{\leq}N/2$. 
Reality of the displacement $u(x,y)$ is guaranteed by requiring
$u(-{\vec{Q}}) = u^{\ast}({\vec{Q}})$.  Also,
note that $u(\vec{Q}=0){\neq}0$ allows the center of gravity to 
fluctuate away from the midplane between the walls.

Using the standard Metropolis algorithm, the simulation attempts to
vary one Fourier amplitude, picked randomly, at a time. 
The initial step sizes, which depend upon $\vec{Q}$, are determined
using a simplified form of the analytic theory \cite{PP92}.
After a certain number of Monte Carlo steps (MCS), step sizes are
adjusted using Dynamically Optimized
Monte-Carlo(DOMC)~\cite{BouzidaKumarSwendsen}. Step size optimization
results in an  
acceptance-rejection ratio of about 1/2, thereby minimizing the
autocorrelation time $\tau$. In practice, because the initial values are
already based on a reasonably good approximation, DOMC
adjustment does not significantly improve the efficiency.

The change in bending energy in Eq.(\ref{RealHamiltonianOneMembrane})
after attempting a step in $u(\vec{Q})$ is $K_c L^2 Q^4 /2$ times
the change in $|u(\vec{Q}|^2$, which requires little
time to compute.  In contrast, calculating the change in
the interaction energy with the walls requires a real space representation
of $u(x,y)$.   However, it is not necessary to use a fast Fourier
transform (FFT) 
routine because the linearity of the Fourier transform requires only
recomputing one Fourier term in order to update $u(x,y)$.
The time this takes is only $O(N^2)$ compared to
$O(N^2 \ln N)$ for a standard FFT routine.  Incremental
addition errors are negligible for the longest runs when double
precision is used; alternatively, one could perform FFT at long
intervals to control such an error. The natural choice is made to
approximate the interaction integral over the membrane by a sum over a
set of equally spaced points $(Li/N, Lj/N)$, with $0{\leq}i,j<N$.

\section{Harmonic Interactions and Finite-size Scaling}
\label{HarmonicPotentialAndScaling}
To test the simulation code and investigate convergence to an
infinite, continuous membrane, it is useful to consider a harmonic
interaction energy.  It is also useful to relate the parameters
in the harmonic potential to those in Eq.(\ref{RealHamiltonianOneMembrane})
by expanding $w_a(u)$ to second order about $u=0$,
\begin{eqnarray}
w_a &=& A{\lambda} \exp(-a/\lambda) \left(1 + \frac{z^2}{2\lambda^2}\right) -\nonumber\\
& & - \frac{H}{12{\pi}a^2}\left(1 + 3\frac{z^2}{\lambda^2}\right),
\label{w_a}
\end{eqnarray}
so that the realistic Eq.(\ref{RealHamiltonianOneMembrane}) then takes
the completely harmonic form
\begin{eqnarray}
W_0 &=& \frac{K_c}{2} \int({\nabla}^2u(r))^{2} d^2r + \nonumber\\
& & \frac{B(a)}{2} \int u^2(r)d^2r + w_0(a)L^2.  
\label{H0}
\end{eqnarray}
where $B = (A/\lambda) e^{-a/\lambda} - H / (2{\pi}a^4)$ and $w_0(a)
= A{\lambda} e^{-a/\lambda} - H / (12{\pi}a^2)$. 
The exact solution (valid for finite $L$ and $N/L$) for this harmonic model is
\begin{equation}
\sigma^2 = \frac{T}{L^2} \sum_{q_x, q_y} \frac{1}{K_c (q_x^2 +
q_y^2)^2 + B},
\label{SigmaSingleHarmonicExact}
\end{equation}
and
\begin{equation}
P = A e^{-a/\lambda} \left[1 + \frac{\sigma^2}{2 \lambda^2}\right].
\label{P}
\end{equation}

Equations (\ref{SigmaSingleHarmonicExact}) and (\ref{P}) are useful in two
ways. First, the harmonic approximation given by Eq.\ref{w_a} is good if
$\sigma{\ll}\lambda$. That provides a test of the correctness of the code,
which is written for the general case of realistic potentials and can
then be applied when $\sigma{\ll}\lambda$.  As
an example, consider a membrane with parameters
$N=4$, $L=700\AA$ and a non-harmonic potential with
$A=1$, $H=100$ ($m_{max}=2$), $\lambda=10\AA$, $K_c=1$, $T=323K$, $a=20\AA$,
where \cite{FN2} gives the units for $A$, $H$ and $K_c$ used in this paper.
The simulation gives $\sigma=0.3394{\pm}0.0004\AA$ and $P =
1.2877{\cdot}10^7 {\pm} 200 erg/cm^3$. In this case,
$\sigma_{exact}=0.33954\AA$, and   
\begin{eqnarray}
P &=& A e^{-a/\lambda} \left[1 + \frac{\sigma^2}{2\lambda^2}\right] - \frac{H}{6{\pi}a^3} \left[1 + 6 \frac{\sigma^2}{\lambda^2}\right] = \nonumber\\
& & 1.28774{\cdot}10^7 erg/cm^3,
\label{P_2HarmonicSingle}
\end{eqnarray}
again showing that simulation results are precise.

The second usage of Eqs. (\ref{SigmaSingleHarmonicExact}) and
(\ref{P}) is to obtain $\sigma$ and $P$ as functions of $N$
and $L$ through the finite sums over $\vec{Q}$.
Simulations are always done with a finite number of Fourier
amplitudes and a finite-sized membrane. However, real membranes are
continuous 
and the relevant size may be larger than $1{\mu}m$. So it is important
to see how the results for finite systems can be used to obtain
quantities for dense ($N{\rightarrow}\infty$) and large
($L{\rightarrow}\infty$, $N/L=const$)
systems. Eqs. (\ref{SigmaSingleHarmonicExact}) and (\ref{P}) can be
used to compute $\sigma(N, L)$ and $P(N, L)$ numerically to examine the
asymptotic behavior of these functions. The result of such analysis
is an asymptotic relation
\begin{equation}
\label{sigma_scaling}
\sigma {\approx} \sigma_\infty - C_1 \left(\frac{L}{N}\right)^2 - C_2
\frac{1}{L^2}, 
\end{equation}
where typically $C_1 \sim 10^{-5} \AA^{-1}$ and $C_2 \sim 10^3 \AA^3$.
The variability caused by the $C_2$ term is very small;
typically about 0.2\% when $L{\geq}700\AA$. However, the $C_1$ term causes
$\sigma$ for a finite membrane to vary with $N$ as much as 20\%.

\section{Obtaining Results for Realistic Interaction Potentials}
\label{RealPot}

\begin{figure}[h]
\begin{center}
\leavevmode
\epsfxsize 6.5cm
\epsffile{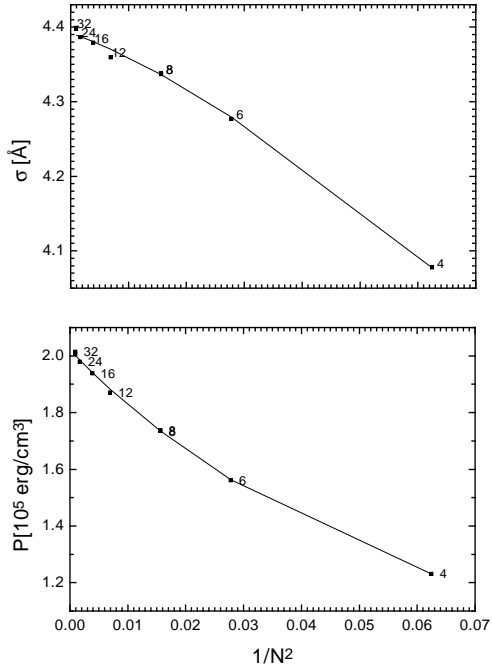}
\caption{$\sigma$ and $P$ vs. $1/N^2$ for $A=1$, $H=0$, $\lambda=1.8\AA$,
$a=20\AA$, $K_c=1$, $T=323K$ and $L=700\AA$.}
\label{sd1pn}
\end{center}
\end{figure}

Table \ref{ScalingResultsSingleRealPotential} shows results for two
selected non-harmonic potentials and a variety of sizes.
One may first note that the
autocorrelation times $\tau_{\sigma^2}$ and $\tau_P$ are nearly
constant with system size. 
Next, convergence with increasing $N$ and constant $L$ is shown in
Fig.\ref{sd1pn} when the vdW interaction is absent.
This behavior is similar to that of a harmonic interaction.
The limiting values can be estimated
by fitting the curve $y = y_\infty + C_2/N^2 + C_3/N^3$. The fits,
shown as solid lines on Fig.\ref{sd1pn}, lead to
$\sigma_\infty = 4.394{\pm}0.004\AA$ and $P_\infty = 202400{\pm}700erg/cm^3$. 
\onecolumn
\begin{table}
\begin{center}
\caption{Representative simulation results for two interactions.}
\label{ScalingResultsSingleRealPotential}
\vspace{0.2 in}
\begin{tabular}
{ c c c c c c c }
N&	$L[\AA]$& $\sigma[\AA]$& $P[\frac{erg}{cm^3}]$& MCS,$10^3$&	$\tau_{\sigma^2}$&$\tau_P$\\\hline
\multicolumn{7} { c }{$A=1$, $H=0$, $K_c=1$ \cite{FN2}, $\lambda=1.8\AA$, $T=323K$, $a=20\AA$} \\\hline
4&	700&	4.0774${\pm}$0.0018&	123010${\pm}$170&	500&	1.59&	1.35\\
6&	700&	4.2767${\pm}$0.0034&	156100${\pm}$400&	100&	1.44&	1.18\\
8&	700&	4.3376${\pm}$0.0028&	173700${\pm}$400&	100&	1.19&	0.96\\
8&	700&	4.3366${\pm}$0.0013&	173470${\pm}$170&	500&	1.21&	0.98\\
12&	700&	4.359${\pm}$0.008&	187000${\pm}$1300&	10&	1.16&	0.97\\
16&	700&	4.3792${\pm}$0.0034&	193800${\pm}$600&	50&	1.08&	0.88\\
24&	700&	4.3864${\pm}$0.0024&	197920${\pm}$430&	30&	0.946&	0.768\\
32&	700&	4.399${\pm}$0.011&	201500${\pm}$1900&	6260&	1.43&	1.41\\
32&	700&	4.3976${\pm}$0.0030&	200600${\pm}$500&	20000&	0.955&	0.741\\\hline
\multicolumn{7} { c }{$A=1$, $H=3$, $m_{max}=4$, $K_c=0.1$, $\lambda=1.4\AA$, $T=323K$, $a=17\AA$} \\\hline
4&	350&	6.0902${\pm}$0.0027&	28000${\pm}$900&	500&	2.46&	1.03\\
6&	525&	6.1097${\pm}$0.0029&	34400${\pm}$900&	200&	2.74&	0.96\\
8&	700&	6.1225${\pm}$0.003&	38500${\pm}$1000&	100&	2.7&	0.97\\
12&	1050&	6.128${\pm}$0.005&	40800${\pm}$1500&	20&	2.73&	1.05\\
16&	1400&	6.1270${\pm}$0.0026&	40000${\pm}$600&	30&	2.35& 	0.86\\
32&	2800&	6.136${\pm}$0.003&	42000${\pm}$600&	6&	2.65&	0.89\\
\end{tabular}
\end{center}
\end{table}
\twocolumn
\vspace{-0.4in}
\begin{figure}[h]
\begin{center}
\leavevmode
\epsfxsize 6.5cm
\epsffile{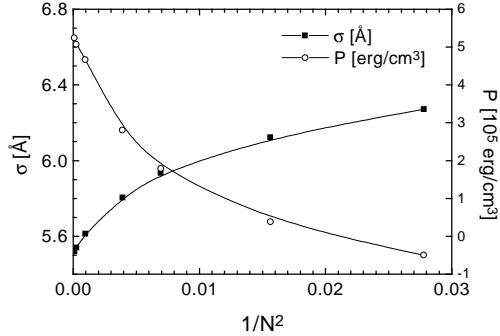}
\caption{$\sigma$ and $P(1/N^2, L=const=700\AA)$ for $A=1$, $H=3$, $m_{max}=4$,
$\lambda=1.4$, $a=17$ and $K_c=0.1$. The lines are drawn to guide the eye.}
\label{sd2nn}
\end{center}
\end{figure} 
Unfortunately, one does not obtain the same asymptotic
behavior as in Fig.\ref{sd1pn}
when the attractive force is large enough that the total potential
has a maximum rather than a minimum 
when in the middle of the space between the walls. For instance,
when $H{\neq}0$, $\sigma$ first decreases with $N$, although later it
gradually levels off and appears to have a minimum.
It is interesting that, while $\sigma$ may change in an unexpected way
as $N$ increases, for the interaction considered, the pressure is
still a smooth quasi-linear function of $1/N^2$
($N{\rightarrow}\infty$), as shown 
in Fig. \ref{sd2nn}, and its limiting value as $N{\rightarrow}\infty$ can
still be estimated by extrapolation.  Despite these variations
in convergence behavior, the associated
changes in $\sigma$ become very small and are certainly less than the desired
accuracy of 1-2\%, so we suggest that it is sufficient to 
increase $N$ only to the point where
further increases result in changes in $\sigma$
and $P$ that are less than the target precision. 

\begin{figure}[h]
\begin{center}
\leavevmode
\epsfxsize 6.2cm
\epsffile{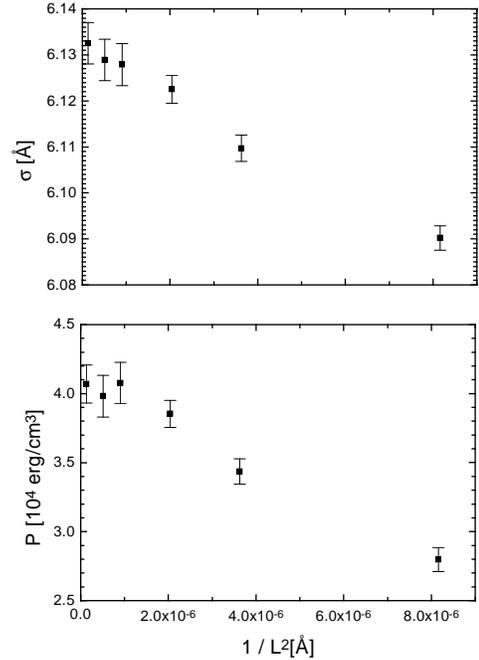}
\caption{$\sigma$ and $P$ vs. $1/L^2$ with $N/L=8/700\AA$ for $A=1$,
$H=3$, $K_c=0.1$ \protect\cite{FN2}, $m_{max}=4$,
$\lambda=1.4$ and $a=17$.}
\label{sd2l}
\end{center}
\end{figure}

The other variable that is potentially significant is the size of the
membrane. Any physical quantity may depend on how large the
membrane is, attaining a certain limiting value as
$L{\rightarrow}\infty$. By increasing $L$ while keeping the
``density'' $N/L = const$, the membrane size is determined
for which $\sigma$ and $P$ approach their limiting
values sufficiently closely. As in the case of harmonic interaction,
the changes in these quantities are relatively small as $L$ is
increased. Indeed, when there is no attractive force, the changes
are so small that they cannot be resolved reliably even when the
estimated statistical errors are of order of $3{\cdot}10^{-3}\AA$. 
When the interaction is smaller, the trends become more pronounced and
similar to those seen for the harmonic potential. An example is given
in Fig.\ref{sd2l} which shows that for a moderate sized membrane the
results approach smoothly and closely those 
for an infinite membrane($L{\rightarrow}\infty$). For $L=700\AA$ the
difference between the estimated limiting value of $\sigma$ and the
observed one 
at $700\AA$ is less than 0.5\%, while for the pressure the same
difference is less 
than 5\% which is about the same as the experimental uncertainty in $P$.

In summary, of the two factors that could affect convergence of
simulation results, i.e. $N$
and $L$, $N$ is most important. $L$ is therefore fixed, typically at
$700\AA$.  $N$ is increased until the changes in quantities of interest
are less than the target precision.  We then fit
a simple function such as $y = y_\infty + c_2/N^2 + c_3/N^3$ to the
sequence of finite $N$ results to estimate $y_\infty$.

\section{Comparison of FMC and Standard PMC Methods}
\label{ComparisonWithReal-Space}

\subsection{Basics of the PMC Simulation Method}
The standard way to simulate membranes~\cite{LipowskyZielinska} will be
called the pointwise MC (PMC) method in which the potential of the
system is given in discretized form
\begin{eqnarray}
W &=& \frac{K_c}{2} \frac{N^2}{L^2} \sum_{ij} (\sum_{nn}u - 4u_{ij})^2 +\nonumber\\
& & \frac{L^2}{N^2} \sum_{ij} w(u_{ij}),
\label{DiscretizedH}
\end{eqnarray}
where $\sum_{nn}u$ is the sum of displacements of nearest neighbors of
site $(i,j)$.  For a harmonic potential,
$w(u) = w_0 + Bu^2/2$, and for periodic boundary conditions
the exact solution for the mean square displacement is 
\begin{eqnarray}
\sigma^2 &=& \frac{k_B T}{L^2} \sum_{\vec{Q}} (B + 4K_C\frac{N^4}{L^4}(\cos(Q_x\frac{L}{N})+ \nonumber\\
& & \cos(Q_y\frac{L}{N})-2)^2 ) ^{-1},
\label{SigmaRealExact}
\end{eqnarray}
where $Q_{x,y} = 2{\pi}n/L$, $-\frac{N}{2}+1{\leq}n{\leq}\frac{N}{2}$.
As with the FMC method, such an exact solution is useful in checking
correctness of the simulation code.

The standard Metropolis algorithm is used, moving one point at a time
in the PMC method.  To start the simulation, an effective B is 
estimated using perturbation theory\cite{PP92}. It is then used in a formula
that gives the mean-square fluctuation of a point (assuming harmonic
potential) about its equilibrium position, determined by its
environment:
\begin{equation}
\sigma_{local} = \sqrt{\frac{k_B T}{B L^2 / N^2 + 20K_c N^2 /L^2}}
\label{SigmaLocalHarmonic}
\end{equation}
Eq.(\ref{SigmaLocalHarmonic}) gives the initial step size. After
a certain number of steps, DOMC\cite{BouzidaKumarSwendsen} is used to
compute the optimal step size, which is used thereafter.
Some results using the PMC method are presented in 
Table \ref{RealSpaceComplexityTable}.

\newpage
\begin{table}
\begin{center}
\caption{Real space simulations of membranes with different density of points,
constrained by a harmonic potential with
$B=8.303{\cdot}10^{11}erg/cm^4$ obtained from $A=1$, $H=0$, $K_c=1$
\protect\cite{FN2}, $\lambda=1.8\AA$, $a=20\AA$. $T=323K$,
$L=700\AA$. Simulation lengths are measured in $10^6$ MCS.}
\label{RealSpaceComplexityTable}
\vspace{0.2 in}
\begin{tabular}
{ c c c c c }
N&	$\sigma[\AA]$&		MCS&	MCS$_{0.1\%}$$^*$&$\tau_{\sigma^2}$ \\\hline
4&	8.390${\pm}$0.005&	1&	0.41&	4.36\\
6&	8.481${\pm}$0.008&	1&	0.98&	13.8\\
8&	8.332${\pm}$0.031&	0.2&	2.77&	41.9\\
8&	8.347${\pm}$0.032&	0.2&	2.94&	42.3\\
8&	8.305${\pm}$0.010&	2&	2.73&	39\\
12&	8.073${\pm}$0.016&	4&	14.9&	203\\
12&	8.070${\pm}$0.015&	4&	14.6&	198\\
16&	8.00${\pm}$0.06&	1&	66&	782\\
16&	8.07${\pm}$0.06&	1&	59&	709\\
\end{tabular}
\end{center}
$^*$ A simulation of approximately such length would have to be done to
attain 0.1\% accuracy for $\sigma$.
\end{table}

\subsection{Comparison of the FMC and PMC methods}
The time required to obtain a target error is one of
the issues determining the viability of any simulation technique. It
is impacted by two separate factors: the relative magnitude of random
errors, and the speed at which various quantities, obtained for a
finite system, converge to their values for the continuous infinite
system. These factors are now considered in detail, to demonstrate the
improvements of the FMC method.
\begin{figure}[h]
\begin{center}
\leavevmode
\epsfxsize 7cm
\epsffile{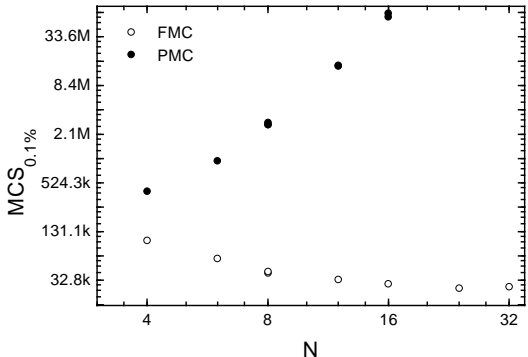}
\caption{Variation with N of the simulation length $MCS_{0.1\%}$,
required for 0.1\% 
precision of $\sigma$, for a PMC simulation of a harmonic potential
with $A=1$, $H=0$, $K_c=1$, $\lambda=1.8\AA$, $a=20\AA$, $T=323K$ and
$L=700\AA$ and for an FMC simulation for a realistic 
model potential with the same parameters.}
\label{length0.1pct}
\end{center}
\end{figure}

The random errors in estimated averages depend on the autocorrelation
times of generated time series. These times are an indication of how
``natural'' the chosen basis is for the simulated system. In the
case of harmonic interactions, the variables used by FMC are exactly
independent and therefore it is possible to vary each of them
separately over its whole range. Although they do become correlated
for realistic interactions, one would still hope that their
dependencies are not great, and so they still represent a good
basis. For PMC simulations, however, the motion of any point is
constrained by its environment, so one would expect the quality of time
series to deteriorate as the ``density'' of the membrane and the
importance of the local environment increase. These assertions are
supported by Tables
\ref{ScalingResultsSingleRealPotential} and \ref{RealSpaceComplexityTable},
which show that for FMC the autocorrelation times remain roughly
constant with increasing $N$, whereas for PMC ${\tau}_{\sigma}$
increases as $N^4$. A related question is how the simulation length
(in MCS) required to obtain a certain accuracy (chosen to be 0.1\%)
varies with N.
A straight line fit to
$\ln(MCS_{0.1\%})$ vs. $\ln N$ dependence for PMC has a slope of
approximately 4 (Fig.\ref{length0.1pct}). Therefore, the amount of time
required to obtain $\sigma$ with the same precision grows as $N^6$ for
PMC method. A somewhat surprising result is that the length required
to achieve a given error 
estimate with FMC decreases with N (Fig.\ref{length0.1pct}). The precise
law governing this decrease is unclear because of the difficulty of
estimating autocorrelation times; one guess, supported by the four points
in the middle ($N=8$ through 24) is that the length
decreases as $1/\sqrt{N}$; however, the hypothesis of the length
staying asymptotically constant cannot be ruled out either. Because 1
MCS (for FMC) takes the amount of time  
$O(N^4)$, the computational complexity of the process generated by a
Fourier-space simulation is only $N^{3.5}$ or $N^4$, assuming that
the same error estimate is achieved. This is a significant improvement
over the $N^6$ law for the real-space simulations.

\begin{figure}[h]
\begin{center}
\leavevmode
\epsfxsize 7cm
\epsffile{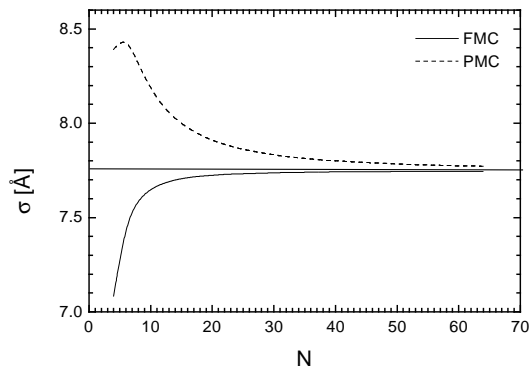}
\caption{Exactly computed $\sigma(N, L=700\AA)$ for Fourier-space
(Eq.(\protect\ref{SigmaSingleHarmonicExact})) and real-space
(Eqs.(\protect\ref{DiscretizedH}) and (\protect\ref{SigmaRealExact}))	
models of a harmonic potential with
$B=8.303{\cdot}10^{11}erg/cm^4$. The other parameters are $K_c=1$,
$T=323K$ and $L=700\AA$.} 
\label{converge_real_fourier}
\end{center}
\end{figure}

The second factor favoring FMC concerns how closely the bending energy
is approximated by the discrete approximation in Eq.(\ref{DiscretizedH}).
This can be evaluated by the exact result for $\sigma$ for a harmonic model.
Fig.\ref{converge_real_fourier} shows that one requires larger $N$ to
obtain the same 
precision with the discrete approximation to the bending energy
required by the PMC method in Eq.(\ref{DiscretizedH})
than for the true continuum model that can be treated naturally by
the FMC method.

A specific example illustrates the preceding principles and also
gives some typical computer times for these simulations.
The example is the harmonic model with parameters given in
Fig.\ref{converge_real_fourier}. 
For the PMC simulation, $N=46$ was chosen so that $\sigma_{exact}(46,
L=700\AA)=7.7898$ was within 0.5\% of its value $7.7478\AA$
for a continuous membrane. A simulation
of 800,000 MCS took 9.5 hours on an SGI workstation with MIPS R5000 1.0
CPU and 128 Mb of RAM, running IRIX 6.2 and resulted in 
${\sigma}= 7.33{\pm}0.19$.  So, 9.5 hours were insufficient to
obtain $\sigma$ with 0.5\% accuracy, and about
$9.5{\cdot}(0.19/(0.005{\cdot}7.75))^2 {\approx} 229$ hours would be
required to achieve that precision.
Turning to FMC, for $N=16$ the exact $\sigma=7.7111\AA$.  
A run of 10,000 MCS yielded $\sigma=7.7184{\pm}0.0165$ and
required only 240 seconds on the same computer as the PMC
simulation. 
One may also compare the time it takes to
obtain the same estimates of random errors for the same $N$ for the
two methods. To do this, $N=16$ and a target error of about 1\% were
chosen for the same interaction as before. A PMC simulation for
300000 MCS took 1174 seconds on an 
SGI workstation with a similar configuration to the one used in the
previous test and resulted in $\sigma = 8.032{\pm}0.082 \AA$ ($\tau_E =
14.7$, $\tau_{\sigma^2} = 441$), a slightly bigger error than
desired. In contrast, an FMC simulation (also with $N=16$) for 2000 MCS
took only 63 seconds on the same computer, and resulted in $\sigma =
7.674{\pm}0.070 \AA$ ($\tau_E = 2.19$, $\tau_{\sigma^2} = 1.44$), the random
error in $\sigma$ now being slightly better than the target. So, in
addition to a much faster convergence of the expected value to one for
a continuous membrane, the FMC method is also the faster one to
obtain a given estimate of stochastic errors.

\section{Results and Implications}
\label{AccurateResults}

\subsection{Distribution of the membrane displacements}
\label{DistributionOfMembraneDisplacements}
\begin{figure}[h]
\begin{center}
\leavevmode
\epsfxsize 7cm
\epsffile{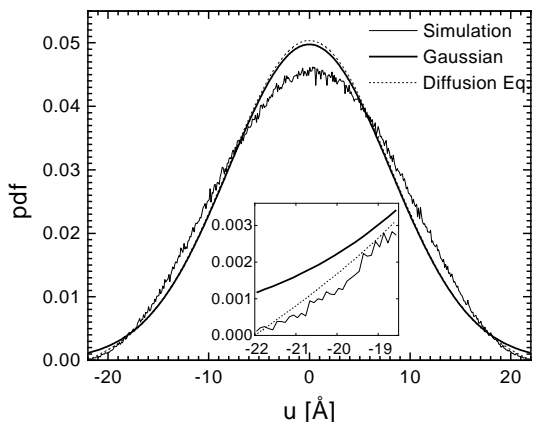}
\caption{Membrane pdf for a realistic constraining
potential. $A=0.2$, $H=0.5$, $\lambda=1.3\AA$, $m_{max}=3$,
$T= 323K$, $K_c=0.1$, $a=22\AA$, $N=32$ and
$L=700\AA$. Also shown are the Gaussian pdf, corresponding to $\sigma
= 8.0196\AA$ and the approximate pdf for the case of pure steric
constraint proposed in \protect\cite{PP92}(Eq.(20)).}
\label{32_700_1_pdf}
\end{center}
\end{figure}

\begin{figure}[h]
\begin{center}
\leavevmode
\epsfxsize 7cm
\epsffile{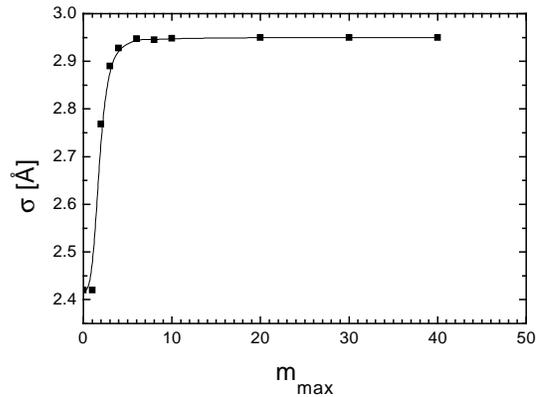}
\caption{The relationship between the number of terms in the expansion
approximating van der Waals potential and $\sigma$, for the parameter
set $a=1$, $H=6$, $\lambda=1.8$, $K_c=0.2$, $T=323$, $a=13$,
$L=700$. The line is drawn to guide the eye.}
\label{mmax_si_sig}
\end{center}
\end{figure}

The functional form of the probability density function (pdf)
is a central assumption in the perturbation theory \cite{PP92}.
Also, the behavior of the pdf near the walls is
significant in discussing the formal divergence of
the van der Waals potential and the importance of the hard
wall collision pressure $P_1$.  If the
pdf does not decay to zero sufficiently quickly
near the walls, then the value of $m_{max}$ used in the power
series expansion would be a sensitive parameter and one
would expect many hard collisions with the walls.
The inset to Fig.\ref{32_700_1_pdf} shows that the
pdf decays to zero near the walls in much the way that
is postulated by theory\cite{PP92}. This is consistent
with our results that  $P_1$ is small and $m_{max}$ is
an insensitive parameter.  
This latter point is explicitly illustrated in Fig.\ref{mmax_si_sig} which
shows that the results for $\sigma$ plateau for $6<m_{max}<40$; a similar
plateau occurs for $P$.  
Finally, Fig.\ref{32_700_1_pdf} shows that, away from the walls,
the pdf is noticeably different from the theoretically assumed
pdf\cite{PP92} and it is generally different from a Gaussian.



\subsection{$P$ and $\sigma$}
\begin{figure}[h]
\begin{center}
\leavevmode
\epsfxsize 7cm
\epsffile{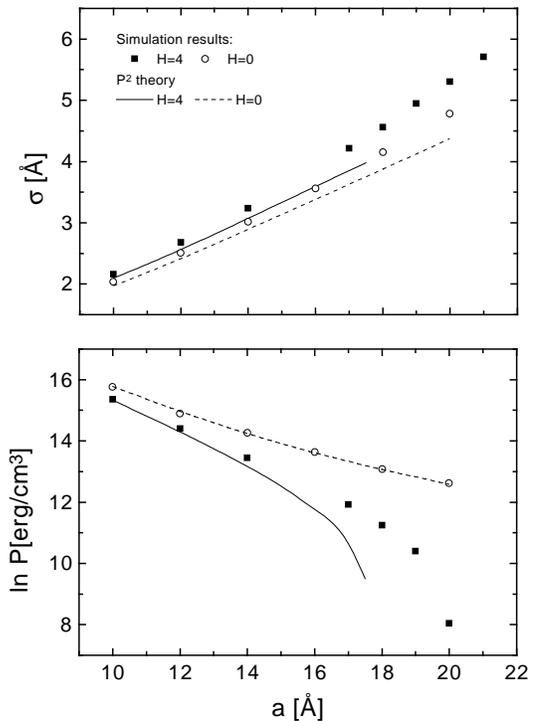}
\caption{$\sigma(a)$ and $\ln P(a)$, obtained from a simulation for
$A=1$, $H=4$, $\lambda=1.8$, $K_c=0.2$ and also for H=0 (all other
parameters being the same) and corresponding results from the
perturbation theory\protect\cite{PP92}.}
\label{sigma_lnp_real}
\end{center}
\end{figure}
For any kind of interaction, the main results to compare to experiment
are the 
relationships between $\ln P$ and $a$, and $\sigma$ and $a$. Figure
\ref{sigma_lnp_real} shows $\ln P$ and $\sigma$ for several values of
$a$. Two interaction types
are considered: $A=1$, $H=4$, $\lambda=1.8$, $K_c=0.2$ and the same
set with $H=0$. These figures also show the results 
obtained from the first-order perturbation theory \cite{PP92}.
The largest differences with the simulations occur at larger $a$ and
when $H$ is non-zero.  In particular, the theory under-predicts
the value of $a$ at $P=0$ when no osmotic pressure is applied.
Overall, however, the theory predicts trends quite well.  

\subsection{Comparison to Experiment}

Recently, it has been proposed that the pressure
due to fluctuations, $P_{fl}$, can be obtained from x-ray line
shape data \cite{HP97}.  The derivation involves the
use of harmonic Caille theory\cite{Cai72,Zhang94}, which yields
\begin{equation}
P_{fl} = - \left ( 
\frac{4}{\pi} \frac{k_B T}{8} 
\right )^2 
\frac{1}{K_c}
\frac{d{\sigma}^{-2}}{da} ,
\label{Pfl}
\end{equation}
where $\sigma$ is obtained from
\begin{equation}
\sigma^2 = {\eta}_1D^2/{\pi}^2 ,
\label{sigma_eta}
\end{equation}
where $\eta_1$ is the Caille parameter determined by the line shape.
The experimental data for three different lipids indicated that
$P_{fl}$ could be represented by an exponential $exp(-a/{\lambda}_{fl})$,
in agreement with the result of perturbation theory \cite{PP92}, but
that ${\lambda}_{fl}$ was significantly greater than $2\lambda$ instead 
of exactly $2\lambda$ given by perturbation theory.  Since neither
the perturbation theory nor the harmonic interpretation of the data
are necessarily correct, it is valuable to test these predictions
using simulations.

\begin{figure}[h]
\begin{center}
\leavevmode
\epsfxsize 7cm
\epsffile{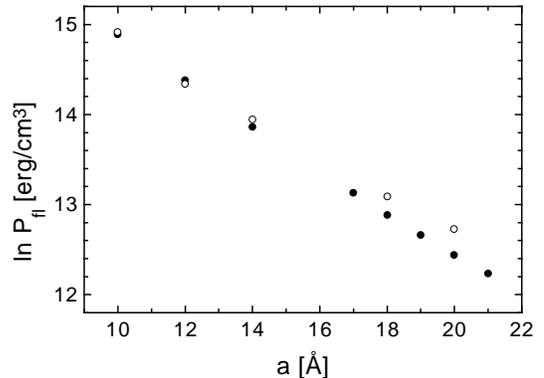}
\caption{Simulation results for $P_{fl}$ vs. $a$ for $A=1$, $H=4$,
$K_c=0.5$ \protect\cite{FN2}, $\lambda=1.8\AA$. Solid circles show
$P_{fl}$ obtained from Eq.(\protect\ref{Psep}) with a slope
${\lambda}_{fl}=4.1\AA$. Open circle show $P_{fl}$ obtained from
Eq.(\protect\ref{Pfl}) with a slope ${\lambda}_{fl}=4.6\AA$.} 
\label{p_fluct_plot}
\end{center}
\end{figure}

Figure \ref{p_fluct_plot} shows two ways of obtaining $P_{fl}$ from
the simulations. The first way uses the definition 
\begin{equation}
P = P_{fl} + P_{b},
\label{Psep}
\end{equation}
where $P$ is the total osmotic pressure and $P_b$ is the pressure with no
fluctuations, {\it i.e.} for the membrane exactly in the middle of
the space between the two walls with $u(x,y)=0$.  The second way uses
Eq.(\ref{Pfl}).  Fig.\ref{p_fluct_plot} shows that the simulated
$P_{fl}$ can be reasonably 
represented by an exponential using either method of computation,
thereby supporting both theory and experiment.  Either method gives
decay lengths ${\lambda}_{fl}$ that exceed $2\lambda$, thereby supporting
experiment.  The two results for $P_{fl}$ in Fig.\ref{p_fluct_plot} do
not, however, 
agree perfectly, and the discrepancy grows for larger
values of $a$.  This is not surprising because the harmonic approximation
is better for small $a$ and progressively breaks down, especially when
the bare potential no longer has a minimum at $z=0$.  This discrepancy
suggests that one should expect some error when subtracting $P_{fl}$ obtained
from Eq.(\ref{Pfl}) from $P$ in Eq.(\ref{Psep}) to obtain $P_b$,
although the error is encouragingly small.  Nevertheless, future work
in this direction can employ simulations to correct this discrepancy and
to allow a better estimate of $P_b$ from which $P_h$, $\lambda$ and $H$
are obtained \cite{HP97}.

\section{Conclusions}
\label{Conclusions}
This paper solves accurately a model of constrained single
membrane fluctuations. The new FMC simulation method provides a way to
simulate accurately, with modest computer resources,
the pressure and mean square fluctuation of a
simple membrane between two hard walls with realistic potentials.
This method is clearly superior to the more conventional
PMC simulation method. Used with typical values of interaction
parameters, it supports the idea of the exponential decay of
fluctuational pressure, lending credibility to a simplified
interpretation of X-ray scattering data in \cite{HP97}.
Finally, the method, with minor modification, may be applied to studies of
more complicated models, such as a stack of membranes or models of
charged lipids and more sophisticated data analysis.

Acknowledgments: We thank Horia Petrache for useful discussions
and acknowledge Prof. R. H. Swendsen for his illuminating expositions of
Monte Carlo technique.  This research was supported by the
U. S. National Institutes of Health Grant GM44976.


\begin{thebibliography}{99}

\bibitem{HP97}
H. I. Petrache, N. Gouliaev, S. Tristram-Nagle, R. Zhang, R. M. Suter,
and J. F. Nagle, submitted to Phys. Rev. E.

\bibitem{Hel78}
W. Helfrich, Z. Naturforsch. 33a, 305 (1978).

\bibitem{Saf89}
C. R. Safinya, E. B. Sirota, D. Roux and G. S. Smith,
Phys. Rev. Lett. 62, 1134 (1989), although a considerable
numerical discrepancy with MC simulations has remained
unresolved, see, e.g., R. R. Netz, Phys. Rev. E 51, 2286
(1995).


\bibitem{SO86}
D. Sornette and N. Ostrowsky, J. Chem. Phys. 84, 4062 (1986).

\bibitem{PP92}
R. Podgornik and V. A. Parsegian, Langmuir 8, 557 (1992).

\bibitem{FN1}
The duration of a
simulation is measured in Monte-Carlo steps (MCS). 1 MCS is defined as
such a sequence of ``moves'' that, on average, changes the variable
corresponding to each degree of freedom once. One MCS is equivalent to
$N^2$ changes of randomly chosen amplitudes for FMC simulations and
for PMC simulations it is equivalent to $N^2$ moves of randomly chosen
points. The auto-correlation times \cite{MullerAndBinder}, denoted
$\tau$ with subscripts referring to physical quantities are also
measured in MCS.

\bibitem{LipowskyZielinska}
R. Lipowsky, B. Zielinska, Phys. Rev. Letts., {\bf{62}}, 1572 (1989)

\bibitem{FNx}
In contrast to the soft confinement regime, extensive simulations
have been performed for single membranes and for short stacks 
in the hard confinement regime using the PMC method.  Some
general reviews include W. Janke, Int. J. Mod. Physics B 4, 1763 (1990),
G. Gompper and M. Schick, Phase Transitions and Critical Phenomena,
Vol. 16 (Academic Press, 1994), eds. C. Domb and J. L. Lebowitz and
R. Lipowsky, Handbook of Biological Physics, Vol. I, Chapter 11
(Elsevier, 1995), eds. R. Lipowsky and E. Sackmann.

\bibitem{Pastor1}
S. E. Feller, R. M. Venable and R. W. Pastor, Langmuir 13, 6555 (1997)

\bibitem{Berko1}
L. Perera, U. Essmann and M. L. Berkowitz,
Progr. Colloid. Polym. Sci. 103, 107 (1997)

\bibitem{Tu1}
K. Tu, D. J. Tobias and M. L. Klein, Biophys. J. 69, 2558 (1995)

\bibitem{Cai72}
A. Caille, C. R. Acad. Sc. (Paris) Serie B 174, 891 (1972).

\bibitem{Als80}
J. Als-Nielsen, J. D. Litster, R. J. Birgeneau, M. Kaplan, C. R. Safinya, A. 
Lindegaard-Anderson and R. Mathiesen, Phys. Rev. B 22, 312 (1980).

\bibitem{Holyst91}
R. Holyst, Phys. Rev. A44, 3692 (1991).

\bibitem{Zhang94}
R. Zhang, R. M. Suter and J. F. Nagle, Phys. Rev. E 50, 5047 (1994).

\bibitem{McI87a}
T. J. McIntosh, A. D. Magid, and S. A. Simon, Biochemistry 26, 7325
(1987).

\bibitem{BouzidaKumarSwendsen}
D. Bouzida, S. Kumar and R. H. Swendsen, Phys. Rev. A, {\bf{45}}, 8894
(1992)

\bibitem{FN2}
In this paper, the following units for the
interaction parameters will be used for brevity: $A[10^9 erg/cm^3]$,
$H[10^{-14} erg]$, $K_c[10^{-12} erg]$.

\bibitem{MullerAndBinder}
H. M\"{u}ller-Krumbhaar and K. Binder, J. Stat. Phys, {\bf{8}}, 1 (1973)

\end{thebibliography}
\end{document}